\documentclass[11pt]{article}

\usepackage{bm}
\usepackage{latexsym}
\usepackage{amssymb}
\usepackage{amsmath}
\usepackage{mathtools}
\usepackage{hyperref}
\hypersetup{colorlinks=true, allcolors=blue}
\usepackage{graphicx}

\usepackage{mathbbol}
\usepackage{braket}
\usepackage{empheq}
\usepackage{cite}
\usepackage[title]{appendix}
\usepackage{here}

\begin{document}
\pagestyle{empty}

\vspace{-4cm}
\begin{center}
    \hfill KEK-QUP-2023-0029 \\
    \hfill KEK-TH-2568 \\
    \hfill KEK-Cosmo-0332 \\
\end{center}

\vspace{2cm}

\begin{center}

{\bf\LARGE  
Quantum entanglement of ions for light dark matter detection
\\
}

\vspace*{1.5cm}
{\large 
Asuka Ito$^{1,2}$, Ryuichiro Kitano$^{2,3}$, Wakutaka Nakano$^{2}$ and Ryoto Takai$^{2,3}$ 
} \\
\vspace*{0.5cm}

{\it 
$^1$International Center for Quantum-field Measurement Systems for Studies of the Universe and Particles (QUP), KEK, Tsukuba 305-0801, Japan\\
$^2$KEK Theory Center, Tsukuba 305-0801,
Japan\\
$^3$The Graduate University for Advanced Studies, SOKENDAI, Tsukuba
305-0801, Japan\\
}

\end{center}

\vspace*{1.0cm}

\begin{abstract}
{\normalsize
\noindent
A detection scheme is explored for light dark matter,
such as axion dark matter or dark photon dark matter,
using a Paul ion trap system.
We first demonstrate that a qubit, constructed from the ground
and first excited states of vibrational modes of ions in a Paul trap,
can serve as an effective sensor for weak electric fields due to its resonant excitation.
As a consequence, a Paul ion trap allows us to search for weak electric fields
induced by light dark matter with masses around the neV range.
Furthermore, we illustrate that an entangled qubit system
involving $N$ ions can enhance the excitation rate by a factor of $N^2$.
The sensitivities of the Paul ion trap system to axion-photon coupling and gauge kinetic mixing
can reach previously unexplored parameter space.
}
\end{abstract} 

%%%%%%%%%%%%%%%%%%%%%%%%%%%%%%%%%%%%%%%%%%%%%%%%%%%%%%%%%%%%%%%%%%%%%%%%%%%%
\newpage
\baselineskip=18pt
\setcounter{page}{2}
\pagestyle{plain}

\setcounter{footnote}{0}

\tableofcontents
\noindent\hrulefill
%%%%%%%%%%%%%%%%%%%%%%%%%%%%%%%%%%%%%%%%%%%%%%%%

\section{Introduction}

Ion traps are innovative technology that utilizes electromagnetic fields
to capture and control ions, in particular, which has been applied
for the implementation of qubits within a quantum computer.
Several trap configurations such as Penning~\cite{gabrielse1984cylindrical,tan1989one}
and Paul traps~\cite{paul1990electromagnetic} have been proposed
and developed toward the achievement of quantum operation with high fidelity.
Interestingly, such quantum operation, like quantum entanglement or squeezing, 
enables us to detect very small electric or magnetic field even beyond
the standard quantum limit.
Therefore, potentially, ion traps can be utilized as quantum sensors
for probing tiny signals from unexplored physics like dark matter.

Dark matter is one of the unsolved puzzles in physics.
The existence of dark matter is supported through the gravitational interaction
by various observations such as
the velocity distribution of the disc galaxies~\cite{Zwicky:1933gu, Zwicky:1937zza},
bullet clusters of galaxies~\cite{Clowe:2003tk, Markevitch:2003at},
the cosmic microwave background~\cite{Planck:2018vyg},
and the large scale structure~\cite{Planck:2018vyg}.
In addition to the gravitational interaction,
dark matter is expected to have interactions with the Standard Model particles.
For instance, the QCD axion was proposed originally by Peccei and Quinn
as the pseudo Nambu--Goldstone boson of broken U(1) symmetry to resolve
the strong CP problem~\cite{Peccei:1977hh, Peccei:1977ur, Weinberg:1977ma, Wilczek:1977pj}.
The QCD axion interacts with photons as
\begin{equation}
    \label{eq:axion_coupling}
    - \frac{1}{4} g_{a \gamma} a F^{\mu\nu} \tilde{F}_{\mu\nu},
\end{equation}
where $a$ denotes the axion,
$F^{\mu\nu}$ is the field strength tensor of the electromagnetic field,
$\tilde{F}_{\mu\nu}$ is its dual, and $g_{a\gamma}$ represents the coupling constant.
Note that, for the QCD axion,
$g_{a\gamma}$ is not an independent parameter but is related to its mass $m_a$:
${g_{a\gamma}}/{m_a} \sim (0.01\text{--}1)~\text{GeV}^{-2}$.
This ratio depends on
the QCD axion models~\cite{Kim:1979if, Shifman:1979if, Zhitnitsky:1980tq, Dine:1981rt}.
One can also consider the axion-like particle (ALP),
which also has the interaction with photons as in Eq.~\eqref{eq:axion_coupling}
with a coupling strength independent from its mass,
while ALPs would not provide a resolution of the strong CP problem.
Both of the QCD axion and ALP can be good dark matter candidates.
In this paper, we will use the term ``axion" to encompass
both QCD axion and ALP for the sake of simplicity.
The dark photon, which is a hidden U(1) gauge field, is also a candidate for dark matter.
It can mix with the Standard Model U(1) gauge field, i.e. photon,
through the kinetic mixing term~\cite{Holdom:1985ag}
\begin{equation}
    \frac{\epsilon}{2} F^{\mu\nu} F'_{\mu\nu}, 
    \label{DP}
\end{equation}
where $\epsilon$ is the mixing parameter 
and $F'_{\mu\nu}$ denotes the field strength tensor of the dark photon.
There have been many works to search for axion and dark photon
through the coupling in Eq.~\eqref{eq:axion_coupling} and in Eq.~\eqref{DP}, respectively,
and they have searched for wide regions in the parameter space.
However, still large parameter space is remained unexplored~\cite{AxionLimits,Caputo:2021eaa}.

In this paper, we explore the possibility of probing axion dark matter and dark photon dark matter
with quantum sensing of Paul traps.
Quantum sensing is known to be useful in searching for light dark matter
such as axion dark matter and dark photon dark matter,
and there have been various recent proposals~\cite{Gilmore:2021qqo, Fan:2022uwu, Chen:2022quj, Afek:2021vjy,
Carney:2021irt, Budker:2021quh, Berlin:2022hfx, Brady:2022bus, Brady:2022qne,
Shi:2022wpf, Sushkov:2023fjw, Ikeda:2021mlv, Chigusa:2023hms, Chigusa:2023szl}.
A trapped ion in a Paul trap has the qubits
consist of its spin states and of its vibrational states, independently.
Utilizing this property, we point out that the ion trap qubit system
enables the enhancement of the signal from light dark matter through qubit operations.
As a consequence, it turns out that Paul traps can
be outstanding quantum sensors for light dark matter detection,
which allows us to probe previously unexplored parameter space of
axion-photon coupling and gauge kinetic mixing.

The axion and dark photon detection by a Penning trap has been
discussed in Ref.~\cite{Gilmore:2021qqo}.
There, two dimensional crystal configurations of ions are used.
It has been discussed that the use of the entanglement
between the vibrational and spin states provides quantum enhancement of the signal.
Our discussion is closely related, but we discuss more on Paul trap configurations
with which various qubit operations have already been realized with a good fidelity.
Which scheme is better depends on technical difficulties in realizing the set-up,
such as boundary conditions of the electric field
as well as the size and strength of the magnetic fields.
We do not try to compare the sensitivities, while we discuss
how quantum manipulations of qubits with high fidelity are useful in the experiments.

This paper is organized as follows.
The basis of a Paul trap system is reviewed briefly in Sec.~\ref{sec:iontrap}.
In Sec.~\ref{sec:dark}, we explain that electric fields
are induced by axion dark matter or dark photon dark matter.
In Sec.~\ref{single}, we show that the small electric field
induced by the light dark matter can be detected with single ion in a Paul trap.
In Sec.~\ref{sec:multi}, the operation for amplifying the signal of the small electric field
caused by the light dark matter in Paul traps with the use of entangled multiple ions is discussed,
and conclusion is given in Sec.~\ref{sec:conclusion}.
In Appendix~\ref{HN}, effects of heating noises on a vibrational qubit is evaluated.
Appendix~\ref{sec:PD} discusses the effect of the decoherence in the $N$-ion set-up.
The potential use of spin qubits, especially in $^{171}$Yb$^{+}$,
instead of vibrational qubits for light dark matter search is studied in Appendix~\ref{sec:Yb}.
%
%The final section is devoted to the conclusion.

%%%%%%%%%%%%%%%%%%%%%%%%%%%%%%%%%%%%%%%%%%%%%%%%%

\section{Overview of trapped ions} \label{sec:iontrap}

Quantum computers utilize qubits for computation, which is similar to how classical computers use bits.
However, they differ significantly in that qubits can exist in superposition states.
This phenomenon is unique to quantum mechanics.
Various physical platforms have been proposed,
such as superconductors, photons, and trapped ions~\cite{ladd2010quantum},
to implement quantum systems with two states that can blend and interact with each other.
Indeed, quantum computers have been realized using techniques
like the Paul trap~\cite{neuhauser1980localized} or the Penning trap~\cite{biercuk2009highfidelity},
following the foundational concept of trapped ion quantum computers proposed in Ref.~\cite{Cirac:1995zz}.

This paper focuses on the Paul trap system.
In this setup, single ionized atoms, such as beryllium, calcium, and ytterbium,
are particularly advantageous due to their ease of ionization.
The electronic states serve as the basis for computational qubits,
referred to as optical, fine structure, hyperfine, and Zeeman qubits,
depending on the selected two levels~\cite{bruzewicz2019trapped}.
We refer to such two level states as a spin qubit regardless of its type for the sake of simplicity, 
and represent the ground and excited states as $\ket{g}$ and $\ket{e}$, respectively.
A distinctive feature of Paul ion traps is the utilization of quantum oscillations of ions.
In the Paul trap system, ions are confined using both static and oscillating
electric fields within an ultra-high vacuum environment.
As a result, a quantum system of coupled harmonic oscillators
is realized by the chain of ions.
The ions are aligned with an approximate interval of $\mathcal{O}(1\text{--}10) \ \mu \text{m}$
by balance between the Coulomb force and the externally applied electric fields.
Let us consider $z$-axis as the alignment direction for the $N$ ions,
along which their oscillations are employed.
Following laser cooling, the ions exhibit behavior resembling quantum harmonic oscillators.
Interactions between ions are facilitated through the Coulomb force,
necessitating the consideration of collective vibrational degrees of freedom.
Specifically, only two states come into play as a qubit:
either all $N$ ions oscillate collectively ($\ket{n = 1}$,
that is a superposition of the first excited state for one of $N$ oscillators)
or they are all at the ground state ($\ket{n = 0}$).
We call the two level states a vibrational qubit in this paper.

Let us illustrate the Paul trap system more specifically.
To this end, we first consider the single ion case for simplicity.
The Hamiltonian is
\begin{equation}
    H_0 = \omega_z a^\dagger a + \frac{\omega_0}{2} \sigma_z.
\end{equation}
Here $\omega_z$ is the angular frequency of the vibrational mode,
$\omega_0$ is the energy gap between the two states $|g\rangle$ and $|e\rangle$,
and $\sigma_z$ is the $z$ component of the Pauli matrix to act on the spin state vector.
Note that usually there is a hierarchy between $\omega_z \sim 10$~MHz and $\omega_0 \sim 10$~GHz.
The four levels $\ket{g,0}$, $\ket{g,1}$, $\ket{e,0}$, and $\ket{e,1}$ are utilized for the implementation.
We set $a^\dagger$ and $a$ as the ladder operators of the vibrational mode
and $\sigma_+$ and $\sigma_-$ as the ladder operators of the spin qubit,
such as $a^\dagger \sigma_- \ket{e,0} = \ket{g,1}$.
The operation controlling the ion states is performed by applying lasers
with frequencies suitable for each ion species.
In the interaction picture,
the general Hamiltonian which describes the interaction between an ion
and a laser with an angular frequency $\omega$ and phase $\phi$ is~\cite{Wineland:1997mg}
\begin{equation}
    H_\text{laser} = \frac{1}{2} \Omega \sigma_+ e^{-i (\omega - \omega_0) t + i \phi}
    + \frac{i}{2} \eta \, \Omega \, a^\dagger \left( \sigma_+ e^{-i (\omega - \omega_0 - \omega_z) t + i \phi}
    - \sigma_- e^{i (\omega - \omega_0 + \omega_z) t - i \phi} \right) + \text{h.c.},
\end{equation}
where $\Omega$ is the Rabi frequency, and $\eta$ is the Lamb-Dicke parameter,
which is typically sufficiently small such as of $\mathcal{O}(0.01\text{--}0.1)$
and thus we omit the higher order terms of $\eta$ in the Hamiltonian~\cite{ozeri2007errors}.
When $\omega = \omega_0$, the first term takes precedence due to resonance, called the carrier resonance.
This phenomenon facilitates the transition of the ion's state between
$\ket{g,n}$ and $\ket{e,n}$ ($n = 0, 1$).
Similarly, lasers with $\omega = \omega_0 - \omega_z$ induce the red sideband resonance.
The red sideband resonance is responsible for the transition between $\ket{g,1}$ and $\ket{e,0}$,
which is described as the third term.
Another condition, $\omega = \omega_0 + \omega_z$, corresponds to the blue sideband resonance.
This resonance involves the mixing between $\ket{g,0}$ and $\ket{e,1}$
and is expressed as the second term.
These three resonances enable all operations on the qubits, 
encompassing not only single qubit gates but also multiple-qubit operations.

Extending the above discussion to multiple ions is straightforward.
The state for a chain of $N$ ions consists of the tensor product of $N$ individual spin qubits
and one vibrational qubit like $\ket{s_1, s_2, \dots, s_N , n}$,
where $s_j = g, e$ and $n = 0, 1$ as explained above.
The fact that all $N$ ions share the vibrational qubit provides the basis for constructing entangled states, as will be discussed in Sec.~\ref{sec:multi}.

%%%%%%%%%%%%%%%%%%%%%%%%%%%%%%%%%%%%%%%%%%%%%%%%

\section{Electric fields induced by light dark matter} \label{sec:dark}

In this section, we explain that both of axion dark matter
and dark photon dark matter can generate electric fields in a Paul trap.
The oscillating electric fields can drive the excitation of the vibrational mode,
and that can be detected by coupling to the spin qubit.
For axion dark matter, the electric field is generated resonantly
through the axion-photon coupling under an external magnetic field~\cite{Sikivie:1983ip}.
For dark photon dark matter, the electric field is produced resonantly through the gauge kinetic mixing.
In both cases, the point is that dark matter around $m_X = \mathcal{O}$(neV) 
behaves as the classical wave since the number of particles
contained in space with the size of the de Broglie wavelength
is much larger than unity.
Moreover, the distribution of dark matter is uniform within the de Broglie wavelength,
which is typically much longer than the size of a Paul trap system.
Thus, one can neglect the spatial dependence of the classical wave.

For the case of axion dark matter, the axion field is given by
\begin{equation}
    a(t) = a_0 \cos (m_a t - \phi_a),
\end{equation}
where it oscillates with the amplitude $a_0$ and the frequency $m_a$.
$\phi_a$ is the phase of the axion dark matter field and is randomly distributed.
The phase does not change during the coherence time
of the dark matter field $T_a = 2 \pi / m_a v_\text{DM}^2$,
where $v_\text{DM} \sim 10^{-3}$ is the relative speed of dark matter
which is determined by the virial velocity of our galaxy~\cite{Ikeda:2021mlv}.
Also, since dark matter is non-relativistic, i.e., $v_\text{DM} \ll 1$,
the amplitude $a_0$ is related to the local dark matter density as
$\rho_\text{DM} = m_a^2 a_0^2 / 2 \sim 0.45~\text{GeV cm}^{-3}$.
In a Paul trap, the system of qubits is under a uniform external magnetic field $\vec{B}$
along the axis direction of a cylinder with a radius of $R$ ($\ll 1 / m_a$)~\cite{PRXQuantum.2.020343}.
Then, axion dark matter can convert to electric field in the presence of the magnetic field.
Taking the direction of the applied magnetic field to be $z$-axis,
the amplitude of electric field induced by axion dark matter is~\cite{Ouellet:2018nfr}
\begin{equation}
    E_{a,z} = \epsilon_a \sqrt{2 \rho_\text{DM}} \sin \left( m_a t - \phi_a \right),  \label{Ea}
\end{equation}
where 
\begin{equation}
    \epsilon_a = \frac{B_z}{2} \frac{g_{a\gamma}}{m_a} (m_a R)^2
    \left[ \left( \log \frac{m_a R}{2} + \gamma - \frac{1}{2} \right)^2
    + \left( \frac{\pi}{2} \right)^2 \right]^{1/2},  \label{epa}
\end{equation}
for $m_a R \ll 1$ and $\gamma = 0.5772 \dots$ is the Euler--Mascheroni constant.

The longitudinal mode of dark photon dark matter with mass of $\mathcal{O}$(neV)
also oscillates in time as the non-relativistic classical field.
The dark electric field component of dark photon dark matter is given by
\begin{equation}
    \vec{E}'(t) = \vec{E}'_{0} \sin \left( m_\text{DP} t - \phi_\text{DP} \right),  \label{E}
\end{equation}
where $m_\text{DP}$ is the mass of dark photon dark matter and $\phi_\text{DP}$ is the random phase.
As in the case of axion dark matter, the amplitude is related to the local dark matter density,
$\rho_\text{DM} =  \vert \vec{E}'_{0} \vert^2 / 2$,
and the coherence time is given by $T_{\rm DP} = 2 \pi / m_{\rm DP} v_\text{DM}^2$.
The dark electric field generates the electric field through the gauge kinetic mixing (\ref{DP}).
Especially, for comparison with Eq.\,(\ref{Ea}),
we pay attention to $z$-component of the electric field induced by dark photon dark matter, that is
\begin{equation}
    E_{\text{DP},z} = \epsilon_\text{DP} \sqrt{2 \rho_\text{DM}}
    \sin \left( m_\text{DP} t - \phi_\text{DP} \right),  \label{Edp}
\end{equation}
where $\epsilon_\text{DP} = \epsilon \cos \theta$.
$\theta \in [0, \pi]$ is the angle between the polarization vector of
the longitudinal mode and the $z$-axis.
In the derivation, we assumed that the vacuum chamber is made of glass
as is usual for linear Paul traps so that one do not need
to put boundary conditions for the electric and magnetic fields,
which usually gives rise to a suppression factor determined by the ratio
between the size of the chamber to the Compton wavelength of dark matter.
On the other hand,
there is the suppression factor of $(m_a R)^2$ in Eq. (\ref{Ea}) in the case of axion dark matter.
This is because axion dark matter can convert to the electric field only in the region of
the existence of an external magnetic field, which is now parameterized by $R$.
However, potentially, we expect that the region where external magnetic field are applied
can be extended independently from the Paul trap system by using such as superconducting magnets.
Then, one may be able to identify $R$ as the size of the magnet, whose typical size is around meters.
Therefore, in the following discussion, we refer to $R$ as the radius of cylindrical magnet.

In the next section, we will show how the electric field induced by the light dark matter 
can be detected with Paul traps.

%%%%%%%%%%%%%%%%%%%%%%%%%%%%%%%%%%%%%%%%%%%%%%%%%

\section{Detecting light dark matter with single ion} \label{single}

The vibrational qubit, introduced in Sec.~\ref{sec:iontrap},
potentially works as a sensor to detect small oscillating electric fields
induced by light dark matter, namely Eqs.~(\ref{Ea}) and (\ref{Edp}).
We demonstrate this by evaluating the excitation rate of the vibrational qubit.

The interaction Hamiltonian between the vibrational qubit and electric field induced by dark matter is given by
\begin{equation}
    \begin{split}
        H_{X} &= \frac{e E_{X,z}}{\sqrt{2 m_\text{ION} \omega_z}}
        (a^\dagger e^{i \omega_z t} + a e^{- i \omega_z t}) \\
        &= e \epsilon_X \sin (m_X t - \phi_X) \sqrt{\frac{\rho_\text{DM}}{m_\text{ION} \omega_z}}
        (a^\dagger e^{i \omega_z t} + a e^{- i \omega_z t}),
    \end{split}
\end{equation}
where $X = a$ or DP represents the case of axion or dark photon and $m_\text{ION}$ is the mass of 
the ion.
This shows that the resonance occurs at $\omega_z = m_X$.
Then, using the rotational wave approximation, 
the time evolution of the vibrational qubit system is described by the displacement operator:
\begin{align}
D(\beta) = \exp \left( \beta a^\dagger - \beta^* a \right) ,  \label{disp}
\end{align}
where
\begin{equation}
    \beta = \alpha_X T, \quad \alpha_X = \frac{e \epsilon_X e^{i \phi_X}}{2} \sqrt{\frac{\rho_\text{DM}}{m_\text{ION} m_X}}.
    \label{eq:beta_X}
\end{equation}
$T$ is time duration of the observation which needs to be 
within the coherence time of light dark matter
$T_X \sim 0.4~{\rm s} \times (10~{\rm neV} / m_X)$, where again $X = a$ or DP.
Starting from $\ket{n = 0}$, this operator develops the vibrational state into 
\begin{equation}
    D(\beta) \ket{0} = \cos \vert \beta \vert \ket{0}
    + \frac{\beta}{\vert \beta \vert} \sin \vert \beta \vert \ket{1}
    \simeq \ket{0} + \alpha_X T \ket{1},
\end{equation}
which gives the probability of the excitation, $\vert \alpha_X \vert^2 T^2$.

The projection into the vibrational eigenstate can be performed by the following sequence.
First, we bring up to the vibtational state into the spin state 
by a red sideband $\pi$ pulse, $c_0 |g,0\rangle + c_1 |g,1 \rangle \to c_0 |g,0\rangle + c_1 |e,0 \rangle$. The spin state is then observed by the protocol of the spin-qubit measurements
 by the observation of the fluorescence after illuminating a laser to bring one of the state into some short-lived state.

In reality, the vibrational qubit is disturbed 
by thermal photons mainly from the electrodes~\cite{RevModPhys.87.1419}.
Taking into account of the excitation of the vibrational qubit by such thermal photons,
as discussed in Appendix \ref{HN}, the total probability of the excitation is given by
\begin{equation}
    P_{1}(T) %= k \vert \alpha_{\rm amp} \vert^2 + \vert \beta \vert^2
    = \dot{\bar{n}} T + \vert \alpha_X \vert^2 T^2 .
    \label{eq:prob_sig}
\end{equation}
The contribution of the noise increases linearly over the duration of measurement $T$,
and the coefficient $\dot{\bar{n}}$ is called a heating rate.

In addition, preparing the ground state of the vibrational qubit cannot be performed perfectly.
Such imperfection is represented by the density matrix $\rho_0$
with the infidelity $\bar{\mathcal{F}}_0$ as
\begin{equation}
    1 - \bar{\mathcal{F}}_0 = \bra{g,0} \rho_0 \ket{g,0}. 
\end{equation}
If the preparation is perfect, the fidelity $1-\bar{\mathcal{F}}_0$ is equal to one.
We further assume that the density matrix of the initial state is depolarized, i.e.,
\begin{equation}
    \rho_0 = \left( 1 - \frac{4}{3} \bar{\mathcal{F}_0} \right) \ket{g,0} \bra{g,0}
    + \frac{\bar{\mathcal{F}_0}}{3} {\bm 1},
\end{equation}
on average during observations.
The cooling time for preparing the initial state is expected to be
$\mathcal{O}(100) \, \mu$s~\cite{PhysRevLett.125.053001}.
Again, the displacement operator describes the time evolution
of the density matrix due to the electric field induced by light dark matter:
\begin{equation}
        \rho_1 = D(\beta) \rho_0 D^\dagger(\beta)
        = \left( 1 - \frac{4}{3} \bar{\mathcal{F}_0} \right)
        \left( \ket{g,0} + \beta T \ket{g,1} \right) \left( \bra{g,0} + \bra{g,1} \beta^* \right)
        + \frac{\bar{\mathcal{F}_0}}{3} {\bm 1}
    \label{rho1}
\end{equation}
within $T_X$.
Here we omitted the heating effect.
One observes the signal of light dark matter by projecting the time evolved density matrix 
onto the excited vibrational qubit state, i.e., $\bra{g,1} \rho_1 \ket{g,1}$.
However, the readout process also contains imperfection, which can be parameterized as
\begin{equation}
    P_{\mathcal{F}}(T) = (1 - \bar{\mathcal{F}}_1) \bra{g,1} \rho_1 \ket{g,1} + (T\text{-independent terms}) , \label{P}
\end{equation}
where $\bar{\mathcal{F}}_1$ represents the infidelity of readout.
The $T$-independent terms are $\mathcal{O}(\bar{\mathcal{F}}_{0,1})$.

From Eqs.\,(\ref{eq:prob_sig}), (\ref{rho1}) and (\ref{P}),
we obtain the readout probability including the heating effect and the infidelity as
\begin{equation}
    P(T) = P(0) + \mathcal{F} (\dot{\bar{n}} T + \vert \alpha_X \vert^2 T^2),
\end{equation}
where $\mathcal{F} = (1 - \mathcal{F}_1) (1 - 4 \mathcal{F}_0 /3)$.
It is reported that the value of infidelity $\bar{\mathcal{F}}_0$ and $\bar{\mathcal{F}}_1$
can reach around $10^{-4}$ for a single ion~\cite{bruzewicz2019trapped}.
Therefore, $\mathcal{F} \approx 1$ and $P(0) \ll \dot{\bar{n}} T$ can be taken for the calculation.

We observe the state after waiting for an observation time $T$ ($<T_X ,\ \dot{\bar{n}}^{-1}$)
and repeat the observation by $N_{\rm shot}$ times,
so that the total time duration of the experiment is $T_{\rm total} \simeq N_{\rm shot} T$.
The number of signals induced by dark matter is given by $S = N_{\rm shot} |\alpha_X|^2 T^2$,
while that due to noise is $B =  N_{\rm shot}\dot{\bar n} T$.
By defining the sensitivity by $S / \sqrt B > 1.645$ (95\% confidence level),
we obtain the sensitivity to the electric field with an angular frequency $\omega_z$ as
\begin{equation}
    E_z = 3.6 \ {\rm nV/m} \times
    \left( \frac{\dot{\bar{n}}}{0.1~\text{s}^{-1}} \right)^{1/4}
    \left( \frac{T_\text{total}}{1~\text{day}} \right)^{-1/4}
    \left( \frac{T}{0.4~\text{s}} \right)^{-1/2}
    \left( \frac{m_\text{ION}}{37~\text{GeV}} \right)^{1/2}
    \left( \frac{\omega_z}{10~\text{neV}} \right)^{1/2}.  \label{Ez}
\end{equation}
Here we assumed the use of $^{40}$Ca$^+$ ion,
and ignored the term of $\log m_a R$ in Eq.~(\ref{epa}),
which just gives an $\mathcal{O}(1)$ factor.
Eq.~(\ref{Ez}) shows that Paul trap systems with a single ion can be good sensors for 
very weak oscillating electric fields at the level of nV/m.
The reference value of the heating rate, $0.1~\text{s}^{-1}$, is
a few times better than the best ones reported in Ref.~\cite{RevModPhys.87.1419}.
Such a low heating rate is required by the condition of $\dot{\bar n} T \ll 1$
if we take the waiting time $T$ to be maximal $T \sim T_X$.
The heating rate is generically smaller for larger trap sizes and lower temperatures.
For an actual experimental set-up, one should optimize the size and temperature
in order to maximize the sensitivity.

By using this sensor, Paul trap systems can be utilized for probing light dark matter, which produce
weak electric fields, namely Eqs.~(\ref{Ea}) and (\ref{Edp}).
Using Eqs.~(\ref{Ea}), (\ref{Edp}) and (\ref{Ez}), sensitivities of a single ion system to
the axion-photon coupling and the gauge kinetic mixing are obtained as follows,
\begin{equation}
    \begin{split}
        g_{a\gamma} &= 4.4 \times 10^{-11}~\text{GeV}^{-1} \times
        \left( \frac{\dot{\bar{n}}}{0.1~\text{s}^{-1}} \right)^{1/4}
        \left( \frac{T_\text{total}}{1~\text{day}} \right)^{-1/4}
        \left( \frac{T}{0.4~\text{s}} \right)^{-1/2}
        \left( \frac{R}{3~\text{m}} \right)^{-2}\\
        &\quad \times \left( \frac{B_z}{100~\text{mT}} \right)^{-1}
        \left( \frac{m_\text{ION}}{37~\text{GeV}} \right)^{1/2}
        \left( \frac{m_a}{10~\text{neV}} \right)^{-1/2}
        \left( \frac{\rho_\text{DM}}{0.45~\text{GeV cm}^{-3}} \right)^{-1/2}
    \end{split}
\end{equation}
and
\begin{equation}
    \begin{split}
        \epsilon &= 6.4 \times 10^{-12} \times
        \left( \frac{\dot{\bar{n}}}{0.1~\text{s}^{-1}} \right)^{1/4}
        \left( \frac{T_\text{total}}{1~\text{day}} \right)^{-1/4}
        \left( \frac{T}{0.4~\text{s}} \right)^{-1/2} \\
        &\quad \times \left( \frac{m_\text{ION}}{37~\text{GeV}} \right)^{1/2}
        \left( \frac{m_\text{DP}}{10~\text{neV}} \right)^{1/2}
        \left( \frac{\rho_\text{DM}}{0.45~\text{GeV cm}^{-3}} \right)^{-1/2}
    \end{split}\label{epcon}
\end{equation}  
at 95\% confidence level, respectively. 
In deriving Eq.\,(\ref{epcon}), we took average on $\theta$.
One can also scan the probed masses of light dark matter
by adjusting the resonance frequency of the vibrational mode.
Potentially, the resonance frequency may be tunable in the range of 100 kHz--10 MHz.
The obtained sensitivities to the axion-photon coupling and the gauge kinetic mixing are depicted
in Fig.~\ref{fig:axion} and Fig.~\ref{fig:darkphoton} in pink, respectively.

%%%%%%%%%%%%%%%%%%%%%%%%%%%%%%%%%%%%%%%%%%%%%%%%%%%%%%%

\section{Quantum enhancement with entangled qubits} \label{sec:multi}

Paul ion traps are designed to be useful for creating quantum entanglement among qubits,
particularly for their implementation in quantum computers.
In this section, we demonstrate that a maximally entangled state of $N$ vibrational qubits,
created through several quantum operations in a Paul trap,
enhances the excitation rate by a factor of $N^2$ compared to the single ion case.

As was explained before, the time evolution of the vibrational qubit caused by light dark matter is 
described by the displacement operator $D(\beta)$ defined as Eq.\,(\ref{disp}).
Here, $\beta = \beta_{\text{r}} + i \beta_{\text{i}}$ is an arbitrary complex number
small enough that its quadratic terms can be ignored.
The displacement operator can be rewritten as
\begin{equation}
    D(\beta) = \left( 1 - i \beta_{\text{r}} \sigma^2_\text{vib} \right) 
       e^{i \beta_{\text{i}} \sigma^1_\text{vib}}
    = \left( 1 + i \beta_{\text{i}} \sigma^1_\text{vib} \right) 
       e^{- i \beta_{\text{r}} \sigma^2_\text{vib}}
\end{equation}
by using the first and second Pauli matrices acting onto vibrational qubits
$\sigma^1_\text{vib} = a^\dagger + a$ and $\sigma^2_\text{vib} = i a^\dagger - i a$.

\begin{figure}[t]
    \centering
    \includegraphics[width=0.6\textwidth]{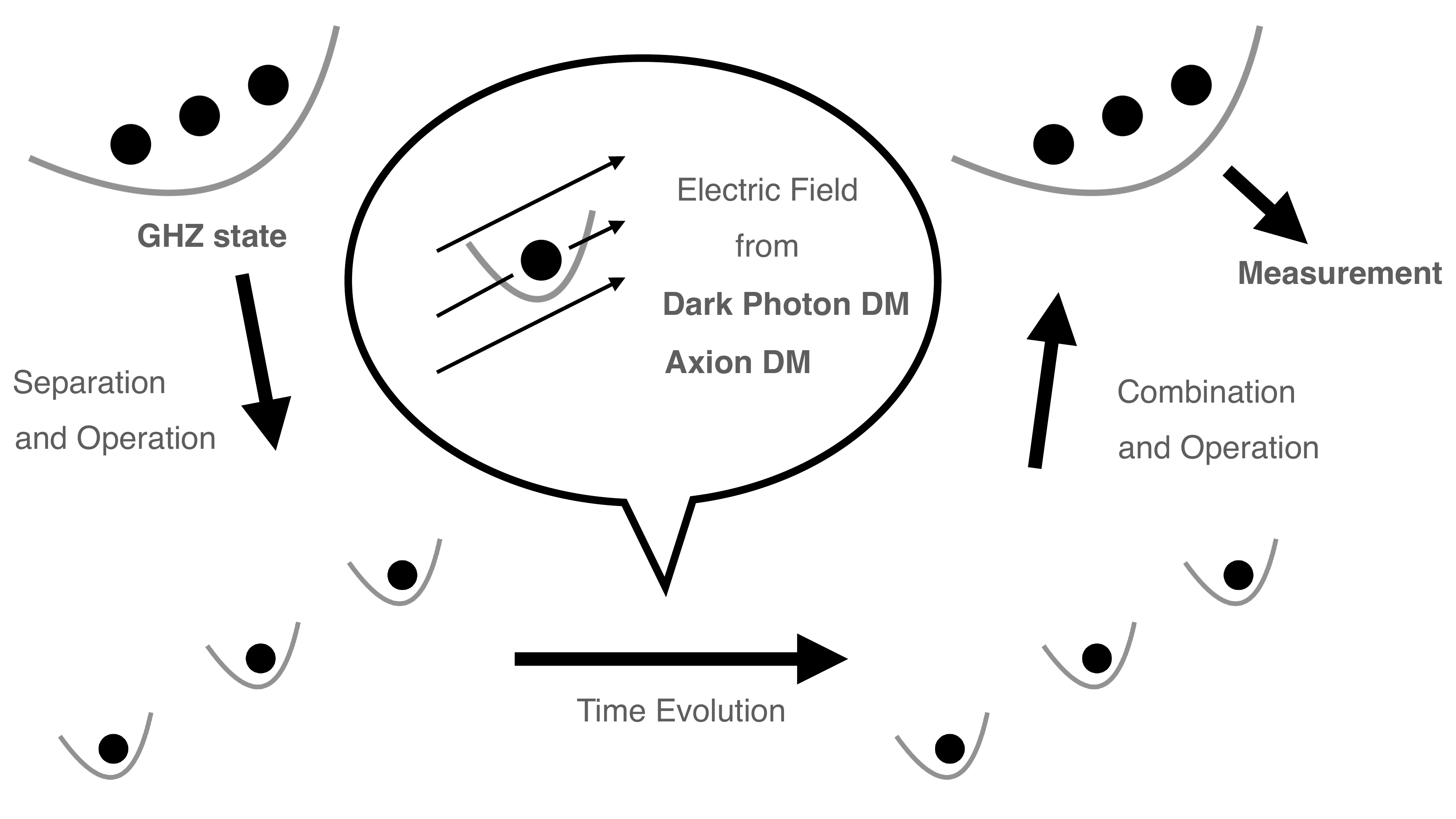}
    \caption{
    The figure illustrates the operational flow for utilizing an entangled $N$-ion system to 
    achieve the $N^2$ enhancement discussed in this section.}
    \label{fig:method}
\end{figure}

We first outline a method for measuring the imaginary part, $\beta_\text{i}$,
whose flow is shown in Fig.~\ref{fig:method}.
The M\o lmer--S\o renson (MS) gate realizes the transformation
\begin{equation}
    \ket{g,g,\dots,g,0} \to |\Psi_{\rm{GHZ}} \rangle =  \frac{1}{\sqrt{2}} \left( \ket{g,g,\dots,g,0} + \ket{e,e,\dots,e,0} \right) \label{GHZ}
\end{equation}
by using the phase rotation operator only onto $\ket{e}$~\cite{molmer1999multiparticle}.
In this state, so-called the Greenberger--Horne--Zeilinger (GHZ) state,
the $N$ spin qubits are maximally entangled.
On the other hand, to access the entanglement of $N$ vibrational qubits,
the ions should be apart from each other so that the each ion system is considered as an isolated system.
After isolating the ions, the ions do not share the vibrational modes any more. We then perform
operations to transform the GHZ state into
\begin{equation}
    \label{eq:state_preparation}
    \ket{\Psi_2} = \frac{1}{\sqrt{2}} \left[ \left( \frac{\ket{g,0} + \ket{g,1}}{\sqrt{2}} \right)^{\otimes N}
    + \left( \frac{\ket{g,0} - \ket{g,1}}{\sqrt{2}} \right)^{\otimes N} \right].
\end{equation}
This is done by acting the Hadamard gate for each ion, 
$\ket{g,0} \to (\ket{g,0} + \ket{e,0}) / \sqrt{2}$
and $\ket{e,0} \to (\ket{g,0} - \ket{e,0}) / \sqrt{2}$, and the red sideband resonance to each ion
to drive $(\ket{g,0} \pm \ket{e,0}) / \sqrt 2 \to (\ket{g,0} \pm \ket{g,1}) / \sqrt 2$.
The electric field induced by light dark matter acts on the state, yielding 
\begin{equation}
    \begin{split}
        \ket{\Psi_3} = \frac{1}{\sqrt{2}} & \left[ e^{i N \beta_\text{i}}
        \left( \frac{\ket{g,0} + \ket{g,1}}{\sqrt{2}}
        + \beta_\text{r} \frac{\ket{g,0} - \ket{g,1}}{\sqrt{2}} \right)^{\otimes N} \right. \\
        & \left. + e^{- i N \beta_\text{i}} \left( \frac{\ket{g,0} - \ket{g,1}}{\sqrt{2}}
        + \beta_\text{r} \frac{\ket{g,0} + \ket{g,1}}{\sqrt{2}} \right)^{\otimes N} \right]. \label{phi3}
    \end{split}
\end{equation}
Since the states $(\ket{g,0} \pm \ket{g,1}) / \sqrt{2}$ 
are the eigenstates of $\sigma^1_\text{vib}$ corresponding to the eigenvalues $\pm 1$,
the new phases $\exp (\pm i N \beta_\text{i})$ appear in each terms.
By applying the inverse operation of the red sideband transition
and the Hadamard gate, one obtains the state
\begin{equation}
    \ket{\Psi_4} = \frac{1}{\sqrt{2}} \left[ e^{i N \beta_\text{i}}
    \big( \ket{g,0} + \beta_\text{r} \ket{e,0} \big)^{\otimes N}
    + e^{- i N \beta_\text{i}} \big( \ket{e,0} + \beta_\text{r} \ket{g,0} \big)^{\otimes N} \right].
    \label{eq:psi4}
\end{equation}
Note that we implicitly assumed that the coherence time of the system
such as the lifetime of the excited states of the qubits, i.e., $\ket{e}$ (usually denoted as $T_1$)
and the decoherence time of the entanglement ($T_2$),
were much longer than the coherence time of light dark matter $T_X$.
We discuss the effect of $T_2$ in Appendix~\ref{sec:PD}.

Since the $\beta_\text{r}$-dependent terms in Eq.~\eqref{eq:psi4}
are orthogonal to the $\beta_\text{r}$-independent terms, we can put them aside.
We now couple all $N$ ions again by bringing them back to the original positions,
so that they share the global vibrational qubit.
The operations of the CNOT (Controlled-NOT) gates
between the first qubit and other qubits $\text{CNOT}_{1,j} \, (j=2,\dots,N)$
bring the component of $\ket{e, e, \dots, e, 0}$ to $\ket{e, g, \dots, g, 0}$
while $\ket{g, \dots, g, 0}$ untouched.
The final operation is the Hadamard gate only for the first qubit
which transfers $(\ket{g, g, \dots, g, 0} \pm \ket{e, g, \dots, g, 0}) / \sqrt{2}$
into $\ket{g, g, \dots, g, 0}$ and $\ket{e, g, \dots, g, 0}$, respectively, so that we obtain
\begin{equation}
    \ket{\Psi_5} \approx \ket{g, g, g, \dots, g, 0} +iN \beta_{\rm i} \, \ket{e, g, g, \dots, g, 0} + \cdots.
\end{equation}
The series of the operations brings back to the original $\ket{g, g, \dots, g, 0}$ state if $\beta = 0$.
The signal of dark matter is the excitation of the first qubit with the probability of
\begin{equation}
   \vert \braket{e, g, g, \dots, g, 0 | \Psi_5} \vert^2 = N^2 \beta_\text{i}^2. \label{N2}
\end{equation}
It is amusing to find that the probability is enhanced by $N^2$
due to the maximal entanglement of the GHZ state.
Similarly, the signal of $\beta_\text{r}$ can be detected with an amplification factor of $N^2$
by following almost the same procedure by replacing $\ket{\Psi_2}$ by
\begin{equation}
    \ket{\Psi_{2'}} = \frac{1}{\sqrt{2}} \left[ \left( \frac{\ket{g,0}
    + i \ket{g,1}}{\sqrt{2}} \right)^{\otimes N}
    + \left( \frac{\ket{g,0} - i \ket{g,1}}{\sqrt{2}} \right)^{\otimes N} \right].
\end{equation}
Note that this state is composed of the eigenstates of $\sigma^2_\text{vib}$.
In our situation the averages of the real and imaginary parts of signal 
are expected to be the same, i.e., $\overline{\beta_{\rm r}^2} =
\overline{\beta_{\rm i}^2} = |\alpha_X|^2 T^2 / 2$.

\begin{figure}[t]
    \centering
    \includegraphics[width=0.95\textwidth]{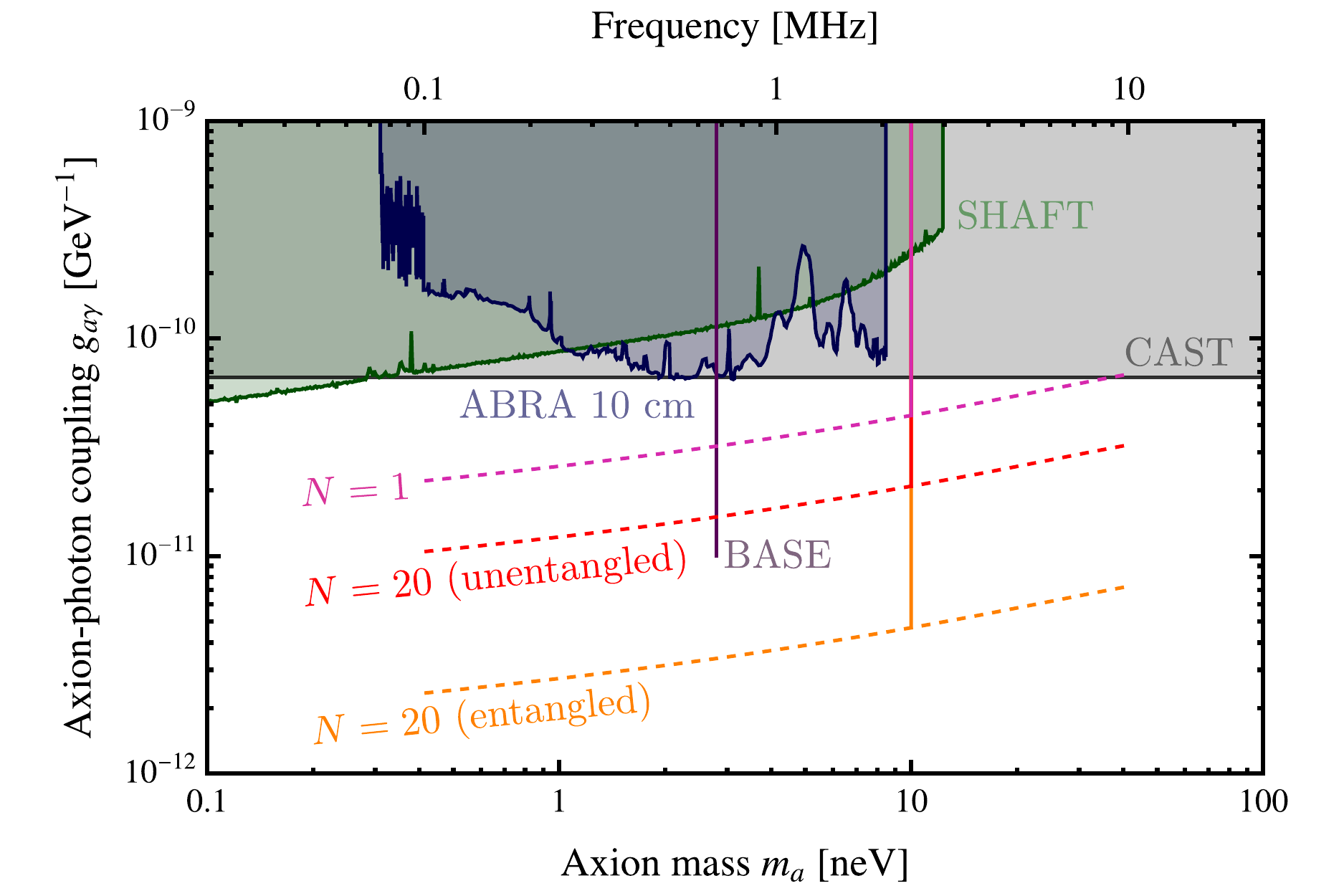}
    \caption[Caption for LOF]{Sensitivities at 95\% C.L.
    to the axion-photon coupling $g_{a\gamma}$ as a function of the axion mass $m_a$ are shown.
    The pink line represents the sensitivity using a single ion.
    The red and orange lines represent the sensitivities 
    using unentangled and entangled twenty ions, respectively.
    The total observation time is set to 1 day and
    the frequency $\omega_z$ is set as 10 neV in the estimation.
    The pink, red, and orange dashed lines show the expected sensitivities for
    the scan of axion masses by adjusting $\omega_z$.
    For comparison, limits by laboratory experiments are shown;
    The black, dark green, blue, and purple regions are excluded 
    by CAST~\cite{CAST:2017uph}, SHAFT~\cite{Gramolin:2020ict}, ABRACADABRA~\cite{Salemi:2021gck},
    and BASE~\cite{Devlin:2021fpq}, respectively\footnotemark.}
    \label{fig:axion}
\end{figure}
\footnotetext{{While there also exists astrophysical constraints on the axion-photon     
    coupling~\cite{Noordhuis:2022ljw,Dessert:2022yqq,Fermi-LAT:2016nkz,Hoof:2022xbe,Escudero:2023vgv},
    they are not shown in the figure because
    the results might be affected by astrophysical uncertainties.}}

The signal enhancement with entangled ions relies on the assumption
that light dark matter excites the vibrational qubits spatially coherently.
This assumption is valid because light dark matter,
whose coherence length is $(m_X v_{\rm DM})^{-1} \sim 100$ km for $m_X = 10$ neV,
homogeneously interact with the ions.
However, whether the heating noise, discussed in Appendix \ref{HN}, disturbs ions coherently or not 
depends on the specific configuration of a Paul ion trap system.
If the heating noise excites ions incoherently, the excitation rate is merely
$N$ times of that of a single ion.
Therefore, the significance of the signal scales as
\begin{equation}
    \frac{S}{\sqrt B} \propto N^{3/2}.
\end{equation}
In the case that the heating noise excites ions coherently,
the excitation rate becomes $N^2$ times of that of a single ion as well as the signal amplification.
Thus, the significance would be $S / \sqrt B \propto N$.
In both cases, it is evident that the use of entanglement is advantageous
in detecting the dark matter signals.

\begin{figure}[t]
    \centering
    \includegraphics[width=0.95\textwidth]{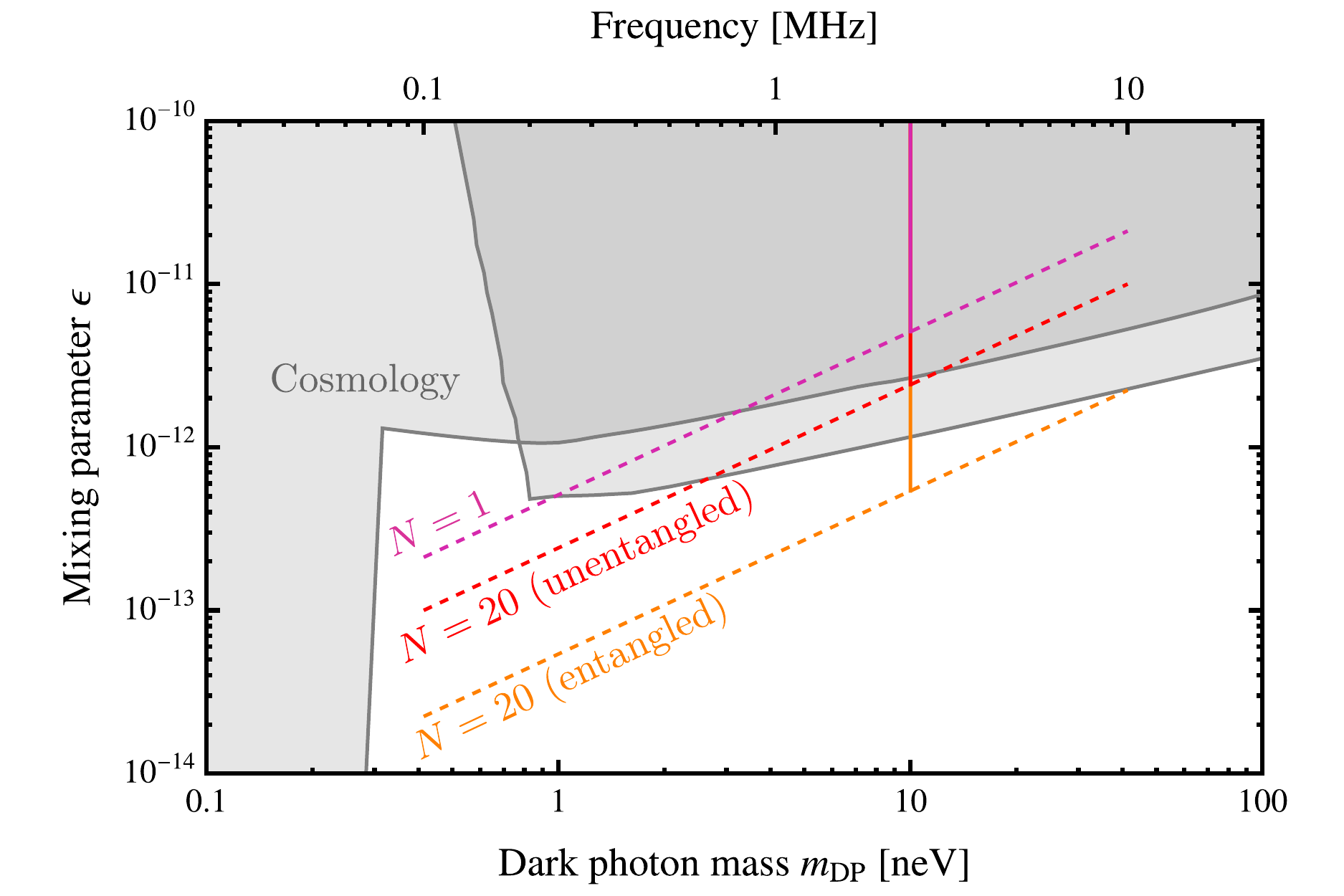}
    \caption{Sensitivities at 95\% C.L.
    to the mixing parameter $\epsilon$ as a function of the dark photon mass $m_\text{DP}$ are shown.
    The pink line represents the sensitivity using a single ion.
    The red and orange lines represent the sensitivities 
    using unentangled and entangled twenty ions, respectively.
    The total observation time is set to 1 day and
    the frequency $\omega_z$ is set as 10 neV in the estimation.
    The pink, red, and orange dashed lines show the expected sensitivities for
    the scan of dark photon masses by adjusting $\omega_z$.
    The two gray regions are excluded by the cosmological observations~\cite{Arias:2012az,Witte:2020rvb}.}
    \label{fig:darkphoton}
\end{figure}

Improved sensitivities to $g_{a\gamma}$ and $\epsilon$ with the use of twenty entangled qubits are depicted 
in Figs.~\ref{fig:axion} and \ref{fig:darkphoton}, respectively.
In the figures, the sensitivity curves depicted in orange represent the case
that ions are entangled and the heating noise interact with them incoherently.
The background due to heating noise may grow up with increasing the number of qubits.
It should be noted that the effect has to be kept small.

One needs to take into account of the fidelity of $N$-ion detector,
which usually tends to get worse compared to the single ion case.
The typical value of the fidelity of GHZ state for multi-ions is found in Ref.~\cite{PRXQuantum.2.020343}.
In the estimation of the sensitivities in Figs.~\ref{fig:axion} and \ref{fig:darkphoton}, 
however, as in the case of the single ion,
we assumed that the infidelity for multiple ions is small enough
compared to the background due to the heating rate. 
Although high fidelity could be achieved for multiple ions in a Paul trap~\cite{PhysRevLett.112.190502}, 
maintaining it for a large number of ions (greater than $\sim$ 20) is challenging.
The situation could change when considering a network of connected Paul traps, with each trap containing
a single ion~\cite{Britton2012,Kielpinski2002,PhysRevLett.124.110501,PhysRevA.89.022317}.

%%%%%%%%%%%%%%%%%%%%%%%%%%%%%%%%%%%%%%%%%%%%%%%%%%%%%%%

\section{Conclusion}
\label{sec:conclusion}

We investigated the potential use of Paul traps as light dark matter detectors.
We first showed that a vibrational qubit of a single ion in the Paul trap
can serve as an effective sensor for electric fields.
Therefore, it can be utilized for probing weak electric fields induced by
axion dark matter and dark photon dark matter with masses in the range of $\mathcal{O}(0.1\text{--}10)$ neV.
Furthermore, in Sec.~\ref{sec:multi}, we demonstrated the scheme for improving the sensitivity
by employing $N$ entangled vibrational qubits to get closer to the Heisenberg limit.
Using this strategy, the number of dark matter signal can be enhanced by a factor of $N^2$.
Consequently, Paul traps can be good quantum detectors for investigating
previously unexplored parameter space of axion-photon coupling and gauge kinetic mixing,
as illustrated in Figs.~\ref{fig:axion} and \ref{fig:darkphoton}.

It should be mentioned that our sensitivity calculations rely on several assumptions.
First, the coherence time of the qubit system parameterized by $T_2$
are assumed to be longer than the coherence time of light dark matter $T_X$.
This assumption would hold true if $T_2 \gtrsim 5$ s
for masses of light dark matter considered in this paper ($\geq 0.1$ neV).
The second assumption is about the infidelity for $N$ multiple ions,
which are assumed to be much smaller than heating noise.
This may remain valid for $N \simeq 20$
even when considering multiple ions in a Paul trap~\cite{PhysRevLett.112.190502},
while it may be difficult for larger $N$.
The situation could change when considering a network connected Paul
traps~\cite{Britton2012,Kielpinski2002,PhysRevLett.124.110501,PhysRevA.89.022317},
potentially enabling high fidelity operations with a large number of ions in the near future.
In particular, ion traps have been extensively studied and developed
for the implementation of qubits within a quantum computer.
It is intriguing to investigate other potential applications of ion traps as proposed in this paper.

%%%%%%%%%%%%%%%%%%%%%%%%%%%%%%%%%%%%%%%%%%%%%%%%%%%%%%%

\section*{Acknowledgements}

We would like to thank Mitsuhiro Yoshida and Shusei Kamioka for discussions.
We also would like to thank Utako Tanaka for providing us with useful information on ion traps.
This work was in part supported by World Premier International
Research Center Initiative (WPI), MEXT, Japan, and JSPS KAKENHI Grant
Numbers JP22K14034~(A.I.), JP23KJ2173~(W.N.), JP22K21350~(R.K.),
JP21H01086~(R.K.), and JP19H00689~(R.K.).

%%%%%%%%%%%%%%%%%%%%%%%%%%%%%%%%%%%%%%%%%%%%%%%%%%%%%%%

\section*{Note added}
While finalizing this manuscript, Ref.~\cite{Chen:2023swh} appeared.
There the authors discuss the possible quantum enhancement of dark matter signals by using
entangled qubits in general quantum computing setup.
The discussion is closely related to the one in Sec.~\ref{sec:multi} of our paper.

%%%%%%%%%%%%%%%%%%%%%%%%%%%%%%%%%%%%%%%%%%%%%%%%%%%%%%%
\appendix
\numberwithin{equation}{section}
\setcounter{equation}{0}
%%%%%%%%%%%%%%%%%%%%%%%%%%%%%%%%%%%%%%%%%%%%%%%%%%%%%%%

\section{Heating noises} \label{HN}

A vibrational qubit is disturbed by thermal photons mainly from the electrodes~\cite{RevModPhys.87.1419}.
In the interaction picture, the interaction Hamiltonian
between the vibrational mode of a single ion and the thermal photons is given by 
\begin{equation}
  H_{\rm heat} = \sum_j g_j  \left( a^{\dagger} b_j e^{-i(\omega_z - \omega(k_j))t} 
                                   + a b_j^{\dagger} e^{i(\omega_z - \omega(k_j))t} \right) ,   \label{heat}
\end{equation}
where $b_j$ ($b_j^{\dagger}$) represents the annihilation (creation) operator
of the thermal photons with a momentum $k_j$, and $g_j$ denotes the coupling strength.
Then, the total Hamiltonian in the presence of light dark matter is
\begin{equation}
  H_{{\rm total}} = \omega_z a^{\dagger} a + \sum_j \omega(k_j) b_j^{\dagger} b_j +  H_X + H_{\rm heat}.
\end{equation}
By using the quantum master equation for the density matrix of the vibrational mode,
which is obtained by tracing out the noise~\cite{walls}, following equations are obtained:
\begin{empheq}[left=\empheqlbrace]{align}
    \frac{d}{dt}\braket{a^{\dagger} a} &= - \gamma \braket{a^{\dagger} a}
                    - i|\alpha_X| e^{i\Delta t} \braket{a^{\dagger}} 
                    +  i|\alpha_X| e^{-i\Delta t} \braket{a} 
                    + \gamma \bar{N}, \nonumber  \\
    \frac{d}{dt}\braket{a}  &=  - i|\alpha_X| e^{i\Delta t} - \frac{\gamma}{2} \braket{a} , \\
    \frac{d}{dt}\braket{a^{\dagger}}  &=   i|\alpha_X| e^{-i\Delta t} - \frac{\gamma}{2} \braket{a^{\dagger}},
                \nonumber\
\end{empheq}
where $\Delta = \omega_z - m_X$ is the detuning between the vibrational mode and the light dark matter.
$\bar{N}$ is the mean particle number of the thermal photons at $\omega_z$ and 
$\gamma = 2\pi \sum_j \delta(\omega_z - \omega(k_j)) g_j^2$.
One can solve the above equations with initial conditions, 
$\braket{a^{\dagger} a}=\braket{a}=\braket{a^{\dagger}}=0$, and obtain
\begin{equation}
  \braket{a^{\dagger} a} =
       - \frac{8|\alpha_X|^2 \cos \Delta t } {4\Delta^2 + \gamma^2} e^{-\gamma t /2}
       +  \frac{4|\alpha_X|^2(1+e^{-\gamma t})}{4\Delta^2 +\gamma^2} 
       + \bar{N}(1-e^{-\gamma t}). \label{nnn}
\end{equation}
When $\gamma t < \Delta t \ll 1$, Eq.(\ref{nnn}) can be approximated to
\begin{equation}
     \braket{a^{\dagger} a} \simeq \gamma \bar{N} t + \vert \alpha_X \vert^2 t^2 . \label{aa}
\end{equation}
One can define a heating rate $\dot{\bar{n}} = \gamma \bar{N}$.
When $\braket{a^{\dagger} a} \ll 1$, Eq.(\ref{aa}) can be regarded as the excitation rate
of the vibrational qubit by the light dark matter.

%%%%%%%%%%%%%%%%%%%%%%%%%%%%%%%%%%%%%%%%%%%%%%%%%%%%%%%

\section{Phase damping} \label{sec:PD}

The interaction between ions and environment causes the decoherence of entangled qubits,
namely, exponential damping of off-diagonal components of density matrix.
This phenomenon is referred to phase damping, and $T_2$ is defined as the lifetime.
It is impossible to remove the phase damping since qubits can interact even with the vacuum environment,
however, it is possible to avoid its effect on signals up to its linear order.
In this Appendix~\ref{sec:PD}, we show that our method to extract maximally enhanced signals, explained
in Sec. \ref{sec:multi},
conducts this mission in long $T_2$ limit.

The effect of phase damping without dissipation can be expressed by the phase kick,
\begin{equation}
    R_z(\theta) = \begin{pmatrix}
    e^{- i \theta /2} & 0 \\
    0 & e^{i \theta /2}
    \end{pmatrix} = \exp \left(- \frac{i \theta}{2} \sigma^3_{\rm vib} \right)
\end{equation}
on the bases $\ket{0}$ and $\ket{1}$,
where $\sigma^3_{\rm vib}$ is the third Pauli matrix onto the vibrational qubit.
The phase $\theta$ is normally distributed with the average zero and the variance $2 \delta$,
which is related to $T_2$ as $\delta \simeq T / T_2$.
In the multi-ion case, each ion $j = 1, \dots, N$ is kicked with individual phase $\theta_j$
according to the same distribution.
As a result, the damping rate of coherence is multiplied up to a factor of $N$.
(If the damping effect acts to all qubits homogeneously, 
the maximum multiplication factor can be $N^2$~\cite{breuer}.)

At the linear order of the signal $\beta$ and the kick $\theta$,
these two effects can be divided.
Starting the maximally entangled state (\ref{eq:state_preparation}),
the density matrix including the time evolution by $\beta$ is
\begin{equation}
    \rho_3 = \ket{\Psi_3} \bra{\Psi_3},
\end{equation}
which is in Eq. (\ref{phi3}).
Additionally the ions are kicked so that their state is changed into
\begin{equation}
    \rho'_3 = \int^{+\infty}_{-\infty} \prod_{j} \left( \frac{{\rm d} \theta_j}{\sqrt{4 \pi \delta}}
    \ e^{- \theta_j^2 / 4 \delta} \right) R_z(\theta) \rho_3 R_z^\dagger(\theta),
\end{equation}
where $R_z(\theta) = R_z(\theta_1) \otimes \cdots \otimes R_z(\theta_N)$.
The state vector $R_z(\theta) \ket{\Psi_3}$ is given as the replacement
$\beta_{\rm r} \to \beta_{\rm r} - i \theta_j / 2$ for the first term
and $\beta_{\rm r} \to \beta_{\rm r} + i \theta_j / 2$ for the second term in Eq. (\ref{phi3}),
with higher order terms neglected.
Following the logic in Sec.~\ref{sec:multi},
these terms do not disturb the state which we collect,
and then the final probability does not change, assuming that $T_2$ is long enough.

There is also phase damping accompanying dissipation originating from factors such as the Hamiltonian (\ref{heat}).
For this factor, the lifetime of coherence would be $T_2 \simeq 2 \dot{\bar{n}}^{-1}$.
(This decoherence disappears at linear order, according to the above discussion.)
Furthermore, for multiple ions, the damping rate of coherence would be multiplied up to a factor of $N$
for a spatially incoherent noise and $N^2$ for a spatially coherent noise, as discussed above.
Therefore, it is essential to mitigate the heating noise in order to realize the procedure explained in 
Sec.~\ref{sec:multi} with high fidelity.
We note that $\dot{\bar{n}}$ and $T_2$ are treated as independent parameters when
estimating sensitivities of entangled qubits to light dark matter
in Fig.~\ref{fig:axion} and Fig.~\ref{fig:darkphoton}.

%%%%%%%%%%%%%%%%%%%%%%%%%%%%%%%%%%%%%%%%%%%%%%%%%%%%%%%

\section{Excitation of spin qubits in \texorpdfstring{$^{171}$Yb$^{+}$}{TEXT}} \label{sec:Yb}

In the main body, we focused only on the vibrational qubits
and demonstrated their ability as dark matter detectors.
For comprehensive study of Paul traps for light dark matter detection,
let us also consider the excitation of spin qubits due to the dark matter.
Especially, we consider $^{171}$Yb$^{+}$ ion whose spin qubit
is composed by hyperfine splitting in $6^2$S$_{1/2}$.
The ground state is the spin singlet state and the excited state is 
one of the spin triplet state as follows,
\begin{equation}
    \ket{g} = \frac{1}{\sqrt{2}} \left( \ket{\uparrow \downarrow} - \ket{\downarrow \uparrow} \right),
    \quad \ket{e} = \frac{1}{\sqrt{2}} \left( \ket{\uparrow \downarrow} + \ket{\downarrow \uparrow} \right),
    \label{Yb}
\end{equation}
where the former and the latter spin symbols represent
those of the outermost electron and the ytterbium nucleus, respectively.
The direct excitation from $\ket{g}$ to $\ket{e}$ can be caused by applying magnetic fields.
Notably, the lifetime of $\ket{e}$ is considerably long $\sim 5\times 10^{11}$ s.
For later convenience, the spin up and down states of an electron and a nucleus
are taken to be the eigenstates of the $z$ component of the spin operators.

In general, axion can also couple to electron field $\psi$ as
\begin{equation}
    \label{eq:axion_electron_coupling}
    \mathcal{L}_{\rm int} = i g_{ae} a \bar{\psi} \gamma_5 \psi ,
\end{equation}
where $g_{ae}$ represents the coupling constant.
Axion dark matter behaves as an effective magnetic field through Eq.\,(\ref{eq:axion_electron_coupling})
by taking the non-relativistic limit~\cite{Barbieri:1985cp}:
\begin{equation}
    \vec{B}_a = \frac{g_{ae}}{e} \vec{v}_\text{DM} \sqrt{2 \rho_\text{DM}}
    \sin \left( m_a t - \phi_a \right),  \label{Ba}
\end{equation}
which is coupling to the spin of electrons.
On the other hand, in addition to the dark electric field (\ref{E}),
the longitudinal mode of dark photon dark matter also has the component of the dark magnetic field.
It induces the magnetic field through the gauge kinetic mixing (\ref{DP}):
\begin{equation}
    \vec{B}_{\text{DP}}= \epsilon \vec{v}_\text{DM} \times \vec{E}'_{0} 
    \sin \left( m_\text{DP} t - \phi_\text{DP} \right).
\end{equation}
Such magnetic fields from the light dark matter can interact with the spin of electrons of ions,
\begin{equation}
        H_{s,X} = \frac{e}{m_e} \vec{B}_X \cdot \vec{S}_e, \label{Hs}
\end{equation}
where $X = a$ or DP.
When the mass of the dark matter coincides with the energy level difference
between $\ket{g}$ and $\ket{e}$, that is 12.6 GHz for $^{171}$Yb$^{+}$ ion,
the dark matter induced magnetic fields resonantly excite the spin qubit through the $z$ component.
At the resonance point, the excitation probability is approximately given by
$(e |\vec{B}_X| T \cos \theta / 4 m_e)^2$, where $\theta$ is the angle 
between the directions of $\vec{B}_X$ and the $z$-axis.

The observation of this excitation is contaminated by the imperfection of 
preparing the ground state and reading out the signal, as discussed in Sec.~\ref{single}.
Following the discussion, the sensitivity of the spin qubit
to the light dark matter can be evaluated as
\begin{equation}
    \begin{split}
        g_{ae} &= 7.9 \times 10^{-3} \times
        \left( \frac{\mathcal{F}}{1} \right)^{-1/2} \left( \frac{P(0)}{10^{-4}} \right)^{1/4}
        \left( \frac{T_\text{total}}{1~\text{day}} \right)^{-1/4} \\
        &\quad  \times
        \left( \frac{T}{8\times 10^{-5}~\text{s}} \right)^{-3/4} 
        \left( \frac{\rho_\text{DM}}{0.45~\text{GeV cm}^{-3}} \right)^{-1/2}
        \left( \frac{v_\text{DM}}{10^{-3}} \right)^{-1}
    \end{split}
\end{equation}
and
\begin{equation}
    \begin{split}
        \epsilon &= 3.2 \times 10^{-2} \times
        \left( \frac{\mathcal{F}}{1} \right)^{-1/2} \left( \frac{P(0)}{10^{-4}} \right)^{1/4}
        \left( \frac{T_\text{total}}{1~\text{day}} \right)^{-1/4} \\
        &\quad  \times
        \left( \frac{T}{8\times 10^{-5}~\text{s}} \right)^{-3/4} 
        \left( \frac{\rho_\text{DM}}{0.45~\text{GeV cm}^{-3}} \right)^{-1/2}
        \left( \frac{v_\text{DM}}{10^{-3}} \right)^{-1}
    \end{split}
\end{equation}
at 95\% confidence level, respectively.
In the derivation we assumed the randomness of the velocity direction and the polarization of dark photon.
One measurement time $T$ is taken to be the coherence time of 
the light dark matter $T_X$, as is done in the main body.
However, it should be noted that this choice is now close 
the typical duration of quantum operations $\sim 100~\mu$s.

It may be useful to give the sensitivity to $B_{X,z}$ ($z$ component of $\vec{B}_X$) as
\begin{equation}
  B_{X,z} = 7.9 \times 10^{-11}\, \text{T} \times \left( \frac{\mathcal{F}}{1} \right)^{-1/2}
                    \left( \frac{P(0)}{10^{-4}} \right)^{1/4}
                   \left( \frac{T_\text{total}}{1~\text{day}} \right)^{-1/4}
                    \left( \frac{T}{8\times 10^{-5}~\text{s}} \right)^{-3/4} .
\end{equation}
$^{171}$Yb$^{+}$ ion can detect oscillating magnetic field of this amplitude at $12.6$ GHz.
If we adopt the setup used in ABRACADABRA~\cite{Kahn:2016aff} where magnetic fields are induced by
the axionic effective current, one can also measure the axion-photon coupling in principle,
although the sensitivity with a single ion would be weaker than SQUID sensors.

As discussed in Sec.~\ref{sec:multi}, we can improve sensitivities with entangled $N$ ions.
For spin qubits, we can use the GHZ state (\ref{GHZ}) as an initial state to be stimulated
by the light dark matter rather than Eq.\,(\ref{phi3}).

%%%%%%%%%%%%%%%%%%%%%%%%%%%%%%%%%%%%%%%%%%%%%%%%%%%%%%%
\bibliography{bibcollection}

\begin{thebibliography}{10}

\bibitem{gabrielse1984cylindrical}
G.~Gabrielse and F.~C. MacKintosh, {\it Cylindrical penning traps with
  orthogonalized anharmonicity compensation,} {\em International Journal of
  Mass Spectrometry and Ion Processes}, vol.~57, no.~1, pp.~1--17, 1984.

\bibitem{tan1989one}
J.~Tan and G.~Gabrielse, {\it One electron in an orthogonalized cylindrical
  penning trap,} {\em Applied physics letters}, vol.~55, no.~20,
  pp.~2144--2146, 1989.

\bibitem{paul1990electromagnetic}
W.~Paul, {\it Electromagnetic traps for charged and neutral particles,} {\em
  Reviews of modern physics}, vol.~62, no.~3, p.~531, 1990.

\bibitem{Zwicky:1933gu}
F.~Zwicky, {\it {Die Rotverschiebung von extragalaktischen Nebeln},} {\em Helv.
  Phys. Acta}, vol.~6, pp.~110--127, 1933,
  \href{http://dx.doi.org/10.1007/s10714-008-0707-4}{{doi:10.1007/s10714-008-0707-4}}.

\bibitem{Zwicky:1937zza}
F.~Zwicky, {\it {On the Masses of Nebulae and of Clusters of Nebulae},} {\em
  Astrophys. J.}, vol.~86, pp.~217--246, 1937,
  \href{http://dx.doi.org/10.1086/143864}{{doi:10.1086/143864}}.

\bibitem{Clowe:2003tk}
D.~Clowe, A.~Gonzalez, and M.~Markevitch, {\it {Weak lensing mass
  reconstruction of the interacting cluster 1E0657-558: Direct evidence for the
  existence of dark matter},} {\em Astrophys. J.}, vol.~604, pp.~596--603,
  2004, \href{http://dx.doi.org/10.1086/381970}{{doi:10.1086/381970}},
  \href{http://arxiv.org/abs/astro-ph/0312273}{{\ttfamily
  arXiv:astro-ph/0312273}}.

\bibitem{Markevitch:2003at}
M.~Markevitch, A.~H. Gonzalez, D.~Clowe, A.~Vikhlinin, L.~David, W.~Forman,
  C.~Jones, S.~Murray, and W.~Tucker, {\it {Direct constraints on the dark
  matter self-interaction cross-section from the merging galaxy cluster
  1E0657-56},} {\em Astrophys. J.}, vol.~606, pp.~819--824, 2004,
  \href{http://dx.doi.org/10.1086/383178}{{doi:10.1086/383178}},
  \href{http://arxiv.org/abs/astro-ph/0309303}{{\ttfamily
  arXiv:astro-ph/0309303}}.

\bibitem{Planck:2018vyg}
N.~Aghanim {\em et~al.}, {\it {Planck 2018 results. VI. Cosmological
  parameters},} {\em Astron. Astrophys.}, vol.~641, p.~A6, 2020,
  \href{http://dx.doi.org/10.1051/0004-6361/201833910}{{doi:10.1051/0004-6361/201833910}},
   \href{http://arxiv.org/abs/1807.06209}{{\ttfamily
  arXiv:1807.06209\,[astro-ph.CO]}}.
\newblock [Erratum: Astron.Astrophys. 652, C4 (2021)].

\bibitem{Peccei:1977hh}
R.~D. Peccei and H.~R. Quinn, {\it {CP Conservation in the Presence of
  Instantons},} {\em Phys. Rev. Lett.}, vol.~38, pp.~1440--1443, 1977,
  \href{http://dx.doi.org/10.1103/PhysRevLett.38.1440}{{doi:10.1103/PhysRevLett.38.1440}}.

\bibitem{Peccei:1977ur}
R.~D. Peccei and H.~R. Quinn, {\it {Constraints Imposed by CP Conservation in
  the Presence of Instantons},} {\em Phys. Rev. D}, vol.~16, pp.~1791--1797,
  1977,
  \href{http://dx.doi.org/10.1103/PhysRevD.16.1791}{{doi:10.1103/PhysRevD.16.1791}}.

\bibitem{Weinberg:1977ma}
S.~Weinberg, {\it {A New Light Boson?},} {\em Phys. Rev. Lett.}, vol.~40,
  pp.~223--226, 1978,
  \href{http://dx.doi.org/10.1103/PhysRevLett.40.223}{{doi:10.1103/PhysRevLett.40.223}}.

\bibitem{Wilczek:1977pj}
F.~Wilczek, {\it {Problem of Strong $P$ and $T$ Invariance in the Presence of
  Instantons},} {\em Phys. Rev. Lett.}, vol.~40, pp.~279--282, 1978,
  \href{http://dx.doi.org/10.1103/PhysRevLett.40.279}{{doi:10.1103/PhysRevLett.40.279}}.

\bibitem{Kim:1979if}
J.~E. Kim, {\it {Weak Interaction Singlet and Strong CP Invariance},} {\em
  Phys. Rev. Lett.}, vol.~43, p.~103, 1979,
  \href{http://dx.doi.org/10.1103/PhysRevLett.43.103}{{doi:10.1103/PhysRevLett.43.103}}.

\bibitem{Shifman:1979if}
M.~A. Shifman, A.~I. Vainshtein, and V.~I. Zakharov, {\it {Can Confinement
  Ensure Natural CP Invariance of Strong Interactions?},} {\em Nucl. Phys. B},
  vol.~166, pp.~493--506, 1980,
  \href{http://dx.doi.org/10.1016/0550-3213(80)90209-6}{{doi:10.1016/0550-3213(80)90209-6}}.

\bibitem{Zhitnitsky:1980tq}
A.~R. Zhitnitsky, {\it {On Possible Suppression of the Axion Hadron
  Interactions. (In Russian)},} {\em Sov. J. Nucl. Phys.}, vol.~31, p.~260,
  1980.

\bibitem{Dine:1981rt}
M.~Dine, W.~Fischler, and M.~Srednicki, {\it {A Simple Solution to the Strong
  CP Problem with a Harmless Axion},} {\em Phys. Lett. B}, vol.~104,
  pp.~199--202, 1981,
  \href{http://dx.doi.org/10.1016/0370-2693(81)90590-6}{{doi:10.1016/0370-2693(81)90590-6}}.

\bibitem{Holdom:1985ag}
B.~Holdom, {\it {Two U(1)'s and Epsilon Charge Shifts},} {\em Phys. Lett. B},
  vol.~166, pp.~196--198, 1986,
  \href{http://dx.doi.org/10.1016/0370-2693(86)91377-8}{{doi:10.1016/0370-2693(86)91377-8}}.

\bibitem{AxionLimits}
C.~O'Hare, ``cajohare/axionlimits: Axionlimits.''
  \url{https://cajohare.github.io/AxionLimits/}, July 2020.

\bibitem{Caputo:2021eaa}
A.~Caputo, A.~J. Millar, C.~A.~J. O'Hare, and E.~Vitagliano, {\it {Dark photon
  limits: A handbook},} {\em Phys. Rev. D}, vol.~104, no.~9, p.~095029, 2021,
  \href{http://dx.doi.org/10.1103/PhysRevD.104.095029}{{doi:10.1103/PhysRevD.104.095029}},
   \href{http://arxiv.org/abs/2105.04565}{{\ttfamily
  arXiv:2105.04565\,[hep-ph]}}.

\bibitem{Gilmore:2021qqo}
K.~A. Gilmore, M.~Affolter, R.~J. Lewis-Swan, D.~Barberena, E.~Jordan, A.~M.
  Rey, and J.~J. Bollinger, {\it {Quantum-enhanced sensing of displacements and
  electric fields with two-dimensional trapped-ion crystals},} {\em Science},
  vol.~373, no.~6555, pp.~673--678, 2021,
  \href{http://dx.doi.org/10.1126/science.abi5226}{{doi:10.1126/science.abi5226}},
   \href{http://arxiv.org/abs/2103.08690}{{\ttfamily
  arXiv:2103.08690\,[quant-ph]}}.

\bibitem{Fan:2022uwu}
X.~Fan, G.~Gabrielse, P.~W. Graham, R.~Harnik, T.~G. Myers, H.~Ramani, B.~A.~D.
  Sukra, S.~S.~Y. Wong, and Y.~Xiao, {\it {One-Electron Quantum Cyclotron as a
  Milli-eV Dark-Photon Detector},} {\em Phys. Rev. Lett.}, vol.~129, no.~26,
  p.~261801, 2022,
  \href{http://dx.doi.org/10.1103/PhysRevLett.129.261801}{{doi:10.1103/PhysRevLett.129.261801}},
   \href{http://arxiv.org/abs/2208.06519}{{\ttfamily
  arXiv:2208.06519\,[hep-ex]}}.

\bibitem{Chen:2022quj}
S.~Chen, H.~Fukuda, T.~Inada, T.~Moroi, T.~Nitta, and T.~Sichanugrist, {\it
  {Detection of hidden photon dark matter using the direct excitation of
  transmon qubits},} 12 2022,
  \href{http://arxiv.org/abs/2212.03884}{{\ttfamily
  arXiv:2212.03884\,[hep-ph]}}.

\bibitem{Afek:2021vjy}
G.~Afek, D.~Carney, and D.~C. Moore, {\it {Coherent Scattering of Low Mass Dark
  Matter from Optically Trapped Sensors},} {\em Phys. Rev. Lett.}, vol.~128,
  no.~10, p.~101301, 2022,
  \href{http://dx.doi.org/10.1103/PhysRevLett.128.101301}{{doi:10.1103/PhysRevLett.128.101301}},
   \href{http://arxiv.org/abs/2111.03597}{{\ttfamily
  arXiv:2111.03597\,[physics.ins-det]}}.

\bibitem{Carney:2021irt}
D.~Carney, H.~H\"affner, D.~C. Moore, and J.~M. Taylor, {\it {Trapped Electrons
  and Ions as Particle Detectors},} {\em Phys. Rev. Lett.}, vol.~127, no.~6,
  p.~061804, 2021,
  \href{http://dx.doi.org/10.1103/PhysRevLett.127.061804}{{doi:10.1103/PhysRevLett.127.061804}},
   \href{http://arxiv.org/abs/2104.05737}{{\ttfamily
  arXiv:2104.05737\,[quant-ph]}}.

\bibitem{Budker:2021quh}
D.~Budker, P.~W. Graham, H.~Ramani, F.~Schmidt-Kaler, C.~Smorra, and S.~Ulmer,
  {\it {Millicharged Dark Matter Detection with Ion Traps},} {\em PRX Quantum},
  vol.~3, no.~1, p.~010330, 2022,
  \href{http://dx.doi.org/10.1103/PRXQuantum.3.010330}{{doi:10.1103/PRXQuantum.3.010330}},
   \href{http://arxiv.org/abs/2108.05283}{{\ttfamily
  arXiv:2108.05283\,[hep-ph]}}.

\bibitem{Berlin:2022hfx}
A.~Berlin {\em et~al.}, {\it {Searches for New Particles, Dark Matter, and
  Gravitational Waves with SRF Cavities},} 3 2022,
  \href{http://arxiv.org/abs/2203.12714}{{\ttfamily
  arXiv:2203.12714\,[hep-ph]}}.

\bibitem{Brady:2022bus}
A.~J. Brady, C.~Gao, R.~Harnik, Z.~Liu, Z.~Zhang, and Q.~Zhuang, {\it
  {Entangled Sensor-Networks for Dark-Matter Searches},} {\em PRX Quantum},
  vol.~3, no.~3, p.~030333, 2022,
  \href{http://dx.doi.org/10.1103/PRXQuantum.3.030333}{{doi:10.1103/PRXQuantum.3.030333}},
   \href{http://arxiv.org/abs/2203.05375}{{\ttfamily
  arXiv:2203.05375\,[quant-ph]}}.

\bibitem{Brady:2022qne}
A.~J. Brady {\em et~al.}, {\it {Entanglement-enhanced optomechanical sensor
  array for dark matter searches},} 10 2022,
  \href{http://arxiv.org/abs/2210.07291}{{\ttfamily
  arXiv:2210.07291\,[quant-ph]}}.

\bibitem{Shi:2022wpf}
H.~Shi and Q.~Zhuang, {\it {Ultimate precision limit of noise sensing and dark
  matter search},} {\em npj Quantum Inf.}, vol.~9, no.~1, p.~27, 2023,
  \href{http://dx.doi.org/10.1038/s41534-023-00693-w}{{doi:10.1038/s41534-023-00693-w}},
   \href{http://arxiv.org/abs/2208.13712}{{\ttfamily
  arXiv:2208.13712\,[quant-ph]}}.

\bibitem{Sushkov:2023fjw}
A.~O. Sushkov, {\it {Quantum Science and the Search for Axion Dark Matter},}
  {\em PRX Quantum}, vol.~4, no.~2, p.~020101, 2023,
  \href{http://dx.doi.org/10.1103/PRXQuantum.4.020101}{{doi:10.1103/PRXQuantum.4.020101}},
   \href{http://arxiv.org/abs/2304.11797}{{\ttfamily
  arXiv:2304.11797\,[hep-ph]}}.

\bibitem{Ikeda:2021mlv}
T.~Ikeda, A.~Ito, K.~Miuchi, J.~Soda, H.~Kurashige, and Y.~Shikano, {\it {Axion
  search with quantum nondemolition detection of magnons},} {\em Phys. Rev. D},
  vol.~105, no.~10, p.~102004, 2022,
  \href{http://dx.doi.org/10.1103/PhysRevD.105.102004}{{doi:10.1103/PhysRevD.105.102004}},
   \href{http://arxiv.org/abs/2102.08764}{{\ttfamily
  arXiv:2102.08764\,[hep-ex]}}.

\bibitem{Chigusa:2023hms}
S.~Chigusa, M.~Hazumi, E.~D. Herbschleb, N.~Mizuochi, and K.~Nakayama, {\it
  {Light Dark Matter Search with Nitrogen-Vacancy Centers in Diamonds},} 2
  2023,  \href{http://arxiv.org/abs/2302.12756}{{\ttfamily
  arXiv:2302.12756\,[hep-ph]}}.

\bibitem{Chigusa:2023szl}
S.~Chigusa, D.~Kondo, H.~Murayama, R.~Okabe, and H.~Sudo, {\it {Axion detection
  via superfluid $^3$He ferromagnetic phase and quantum measurement
  techniques},} 9 2023,  \href{http://arxiv.org/abs/2309.09160}{{\ttfamily
  arXiv:2309.09160\,[hep-ph]}}.

\bibitem{ladd2010quantum}
T.~D. Ladd, F.~Jelezko, R.~Laflamme, Y.~Nakamura, C.~Monroe, and J.~L. O'Brien,
  {\it Quantum computers,} {\em nature}, vol.~464, no.~7285, pp.~45--53, 2010,
  \href{http://arxiv.org/abs/1009.2267}{{\ttfamily
  arXiv:1009.2267\,[quant-ph]}}.

\bibitem{neuhauser1980localized}
W.~Neuhauser, M.~Hohenstatt, P.~Toschek, and H.~Dehmelt, {\it Localized visible
  ba+ mono-ion oscillator,} {\em Physical Review A}, vol.~22, no.~3, p.~1137,
  1980.

\bibitem{biercuk2009highfidelity}
M.~J. Biercuk, H.~Uys, A.~P. VanDevender, N.~Shiga, W.~M. Itano, and J.~J.
  Bollinger, {\it High-fidelity quantum control using ion crystals in a penning
  trap,} 2009,  \href{http://arxiv.org/abs/0906.0398}{{\ttfamily
  arXiv:0906.0398\,[quant-ph]}}.

\bibitem{Cirac:1995zz}
J.~I. Cirac and P.~Zoller, {\it {Quantum Computations with Cold Trapped Ions},}
  {\em Phys. Rev. Lett.}, vol.~74, pp.~4091--4094, 1995,
  \href{http://dx.doi.org/10.1103/PhysRevLett.74.4091}{{doi:10.1103/PhysRevLett.74.4091}}.

\bibitem{bruzewicz2019trapped}
C.~D. Bruzewicz, J.~Chiaverini, R.~McConnell, and J.~M. Sage, {\it Trapped-ion
  quantum computing: Progress and challenges,} {\em Applied Physics Reviews},
  vol.~6, no.~2, p.~021314, 2019,
  \href{http://arxiv.org/abs/1904.04178}{{\ttfamily
  arXiv:1904.04178\,[quant-ph]}}.

\bibitem{Wineland:1997mg}
D.~J. Wineland, C.~Monroe, W.~M. Itano, D.~Leibfried, B.~E. King, and D.~M.
  Meekhof, {\it {Experimental issues in coherent quantum state manipulation of
  trapped atomic ions},} {\em J. Res. Natl. Inst. Stand. Tech.}, vol.~103,
  p.~259, 1998,  \href{http://arxiv.org/abs/quant-ph/9710025}{{\ttfamily
  arXiv:quant-ph/9710025}}.

\bibitem{ozeri2007errors}
R.~Ozeri, W.~M. Itano, R.~Blakestad, J.~Britton, J.~Chiaverini, J.~D. Jost,
  C.~Langer, D.~Leibfried, R.~Reichle, S.~Seidelin, {\em et~al.}, {\it Errors
  in trapped-ion quantum gates due to spontaneous photon scattering,} {\em
  Physical Review A}, vol.~75, no.~4, p.~042329, 2007.

\bibitem{Sikivie:1983ip}
P.~Sikivie, {\it {Experimental Tests of the Invisible Axion},} {\em Phys. Rev.
  Lett.}, vol.~51, pp.~1415--1417, 1983,
  \href{http://dx.doi.org/10.1103/PhysRevLett.51.1415}{{doi:10.1103/PhysRevLett.51.1415}}.
\newblock [Erratum: Phys.Rev.Lett. 52, 695 (1984)].

\bibitem{PRXQuantum.2.020343}
I.~Pogorelov, T.~Feldker, C.~D. Marciniak, L.~Postler, G.~Jacob,
  O.~Krieglsteiner, V.~Podlesnic, M.~Meth, V.~Negnevitsky, M.~Stadler,
  B.~H\"ofer, C.~W\"achter, K.~Lakhmanskiy, R.~Blatt, P.~Schindler, and
  T.~Monz, {\it Compact ion-trap quantum computing demonstrator,} {\em PRX
  Quantum}, vol.~2, p.~020343, Jun 2021,
  \href{http://dx.doi.org/10.1103/PRXQuantum.2.020343}{{doi:10.1103/PRXQuantum.2.020343}}.

\bibitem{Ouellet:2018nfr}
J.~Ouellet and Z.~Bogorad, {\it {Solutions to Axion Electrodynamics in Various
  Geometries},} {\em Phys. Rev. D}, vol.~99, no.~5, p.~055010, 2019,
  \href{http://dx.doi.org/10.1103/PhysRevD.99.055010}{{doi:10.1103/PhysRevD.99.055010}},
   \href{http://arxiv.org/abs/1809.10709}{{\ttfamily
  arXiv:1809.10709\,[hep-ph]}}.

\bibitem{RevModPhys.87.1419}
M.~Brownnutt, M.~Kumph, P.~Rabl, and R.~Blatt, {\it Ion-trap measurements of
  electric-field noise near surfaces,} {\em Rev. Mod. Phys.}, vol.~87,
  pp.~1419--1482, Dec 2015,
  \href{http://dx.doi.org/10.1103/RevModPhys.87.1419}{{doi:10.1103/RevModPhys.87.1419}}.

\bibitem{PhysRevLett.125.053001}
L.~Feng, W.~L. Tan, A.~De, A.~Menon, A.~Chu, G.~Pagano, and C.~Monroe, {\it
  Efficient ground-state cooling of large trapped-ion chains with an
  electromagnetically-induced-transparency tripod scheme,} {\em Phys. Rev.
  Lett.}, vol.~125, p.~053001, Jul 2020,
  \href{http://dx.doi.org/10.1103/PhysRevLett.125.053001}{{doi:10.1103/PhysRevLett.125.053001}}.

\bibitem{molmer1999multiparticle}
K.~M{\o}lmer and A.~S{\o}rensen, {\it Multiparticle entanglement of hot trapped
  ions,} {\em Physical Review Letters}, vol.~82, no.~9, p.~1835, 1999,
  \href{http://arxiv.org/abs/quant-ph/9810040}{{\ttfamily
  arXiv:quant-ph/9810040}}.

\bibitem{CAST:2017uph}
V.~Anastassopoulos {\em et~al.}, {\it {New CAST Limit on the Axion-Photon
  Interaction},} {\em Nature Phys.}, vol.~13, pp.~584--590, 2017,
  \href{http://dx.doi.org/10.1038/nphys4109}{{doi:10.1038/nphys4109}},
  \href{http://arxiv.org/abs/1705.02290}{{\ttfamily
  arXiv:1705.02290\,[hep-ex]}}.

\bibitem{Gramolin:2020ict}
A.~V. Gramolin, D.~Aybas, D.~Johnson, J.~Adam, and A.~O. Sushkov, {\it {Search
  for axion-like dark matter with ferromagnets},} {\em Nature Phys.}, vol.~17,
  no.~1, pp.~79--84, 2021,
  \href{http://dx.doi.org/10.1038/s41567-020-1006-6}{{doi:10.1038/s41567-020-1006-6}},
   \href{http://arxiv.org/abs/2003.03348}{{\ttfamily
  arXiv:2003.03348\,[hep-ex]}}.

\bibitem{Salemi:2021gck}
C.~P. Salemi {\em et~al.}, {\it {Search for Low-Mass Axion Dark Matter with
  ABRACADABRA-10~cm},} {\em Phys. Rev. Lett.}, vol.~127, no.~8, p.~081801,
  2021,
  \href{http://dx.doi.org/10.1103/PhysRevLett.127.081801}{{doi:10.1103/PhysRevLett.127.081801}},
   \href{http://arxiv.org/abs/2102.06722}{{\ttfamily
  arXiv:2102.06722\,[hep-ex]}}.

\bibitem{Devlin:2021fpq}
J.~A. Devlin {\em et~al.}, {\it {Constraints on the Coupling between Axionlike
  Dark Matter and Photons Using an Antiproton Superconducting Tuned Detection
  Circuit in a Cryogenic Penning Trap},} {\em Phys. Rev. Lett.}, vol.~126,
  no.~4, p.~041301, 2021,
  \href{http://dx.doi.org/10.1103/PhysRevLett.126.041301}{{doi:10.1103/PhysRevLett.126.041301}},
   \href{http://arxiv.org/abs/2101.11290}{{\ttfamily
  arXiv:2101.11290\,[astro-ph.CO]}}.

\bibitem{Noordhuis:2022ljw}
D.~Noordhuis, A.~Prabhu, S.~J. Witte, A.~Y. Chen, F.~Cruz, and C.~Weniger, {\it
  {Novel Constraints on Axions Produced in Pulsar Polar-Cap Cascades},} 9 2022,
   \href{http://arxiv.org/abs/2209.09917}{{\ttfamily
  arXiv:2209.09917\,[hep-ph]}}.

\bibitem{Dessert:2022yqq}
C.~Dessert, D.~Dunsky, and B.~R. Safdi, {\it {Upper limit on the axion-photon
  coupling from magnetic white dwarf polarization},} {\em Phys. Rev. D},
  vol.~105, no.~10, p.~103034, 2022,
  \href{http://dx.doi.org/10.1103/PhysRevD.105.103034}{{doi:10.1103/PhysRevD.105.103034}},
   \href{http://arxiv.org/abs/2203.04319}{{\ttfamily
  arXiv:2203.04319\,[hep-ph]}}.

\bibitem{Fermi-LAT:2016nkz}
M.~Ajello {\em et~al.}, {\it {Search for Spectral Irregularities due to
  Photon\textendash{}Axionlike-Particle Oscillations with the Fermi Large Area
  Telescope},} {\em Phys. Rev. Lett.}, vol.~116, no.~16, p.~161101, 2016,
  \href{http://dx.doi.org/10.1103/PhysRevLett.116.161101}{{doi:10.1103/PhysRevLett.116.161101}},
   \href{http://arxiv.org/abs/1603.06978}{{\ttfamily
  arXiv:1603.06978\,[astro-ph.HE]}}.

\bibitem{Hoof:2022xbe}
S.~Hoof and L.~Schulz, {\it {Updated constraints on axion-like particles from
  temporal information in supernova SN1987A gamma-ray data},} {\em JCAP},
  vol.~03, p.~054, 2023,
  \href{http://dx.doi.org/10.1088/1475-7516/2023/03/054}{{doi:10.1088/1475-7516/2023/03/054}},
   \href{http://arxiv.org/abs/2212.09764}{{\ttfamily
  arXiv:2212.09764\,[hep-ph]}}.

\bibitem{Escudero:2023vgv}
M.~Escudero, C.~K. Pooni, M.~Fairbairn, D.~Blas, X.~Du, and D.~J.~E. Marsh,
  {\it {Axion Star Explosions: A New Source for Axion Indirect Detection},} 2
  2023,  \href{http://arxiv.org/abs/2302.10206}{{\ttfamily
  arXiv:2302.10206\,[hep-ph]}}.

\bibitem{Arias:2012az}
P.~Arias, D.~Cadamuro, M.~Goodsell, J.~Jaeckel, J.~Redondo, and A.~Ringwald,
  {\it {WISPy Cold Dark Matter},} {\em JCAP}, vol.~06, p.~013, 2012,
  \href{http://dx.doi.org/10.1088/1475-7516/2012/06/013}{{doi:10.1088/1475-7516/2012/06/013}},
   \href{http://arxiv.org/abs/1201.5902}{{\ttfamily
  arXiv:1201.5902\,[hep-ph]}}.

\bibitem{Witte:2020rvb}
S.~J. Witte, S.~Rosauro-Alcaraz, S.~D. McDermott, and V.~Poulin, {\it {Dark
  photon dark matter in the presence of inhomogeneous structure},} {\em JHEP},
  vol.~06, p.~132, 2020,
  \href{http://dx.doi.org/10.1007/JHEP06(2020)132}{{doi:10.1007/JHEP06(2020)132}},
   \href{http://arxiv.org/abs/2003.13698}{{\ttfamily
  arXiv:2003.13698\,[astro-ph.CO]}}.

\bibitem{PhysRevLett.112.190502}
T.~Choi, S.~Debnath, T.~A. Manning, C.~Figgatt, Z.-X. Gong, L.-M. Duan, and
  C.~Monroe, {\it Optimal quantum control of multimode couplings between
  trapped ion qubits for scalable entanglement,} {\em Phys. Rev. Lett.},
  vol.~112, p.~190502, May 2014,
  \href{http://dx.doi.org/10.1103/PhysRevLett.112.190502}{{doi:10.1103/PhysRevLett.112.190502}}.

\bibitem{Britton2012}
Britton {\em et~al.}, {\it Engineered two-dimensional ising interactions in a
  trapped-ion quantum simulator with hundreds of spins,} {\em Nature},
  vol.~484, no.~7395, pp.~489--492, 2012,
  \href{http://dx.doi.org/10.1038/nature10981}{{doi:10.1038/nature10981}}.

\bibitem{Kielpinski2002}
D.~Kielpinski {\em et~al.}, {\it Architecture for a large-scale ion-trap
  quantum computer,} {\em Nature}, vol.~417, no.~6890, pp.~709--711, 2002,
  \href{http://dx.doi.org/10.1038/nature00784}{{doi:10.1038/nature00784}}.

\bibitem{PhysRevLett.124.110501}
L.~J. Stephenson {\em et~al.}, {\it High-rate, high-fidelity entanglement of
  qubits across an elementary quantum network,} {\em Phys. Rev. Lett.},
  vol.~124, p.~110501, Mar 2020,
  \href{http://dx.doi.org/10.1103/PhysRevLett.124.110501}{{doi:10.1103/PhysRevLett.124.110501}}.

\bibitem{PhysRevA.89.022317}
C.~Monroe, R.~Raussendorf, A.~Ruthven, K.~R. Brown, P.~Maunz, L.-M. Duan, and
  J.~Kim, {\it Large-scale modular quantum-computer architecture with atomic
  memory and photonic interconnects,} {\em Phys. Rev. A}, vol.~89, p.~022317,
  Feb 2014,
  \href{http://dx.doi.org/10.1103/PhysRevA.89.022317}{{doi:10.1103/PhysRevA.89.022317}}.

\bibitem{Chen:2023swh}
S.~Chen, H.~Fukuda, T.~Inada, T.~Moroi, T.~Nitta, and T.~Sichanugrist, {\it
  {Quantum Enhancement in Dark Matter Detection with Quantum Computation},} 11
  2023,  \href{http://arxiv.org/abs/2311.10413}{{\ttfamily
  arXiv:2311.10413\,[hep-ph]}}.

\bibitem{walls}
D.~F. Walls and G.~J. Milburn, {\em {Quantum Optics}}.
\newblock Springer, 1994.

\bibitem{breuer}
H.-P. Breuer and F.~Petruccione, {\em {The Theory of Open Quantum Systems}}.
\newblock Oxford University Press, 2007.

\bibitem{Barbieri:1985cp}
R.~Barbieri, M.~Cerdonio, G.~Fiorentini, and S.~Vitale, {\it {AXION TO MAGNON
  CONVERSION: A SCHEME FOR THE DETECTION OF GALACTIC AXIONS},} {\em Phys. Lett.
  B}, vol.~226, pp.~357--360, 1989,
  \href{http://dx.doi.org/10.1016/0370-2693(89)91209-4}{{doi:10.1016/0370-2693(89)91209-4}}.

\bibitem{Kahn:2016aff}
Y.~Kahn, B.~R. Safdi, and J.~Thaler, {\it {Broadband and Resonant Approaches to
  Axion Dark Matter Detection},} {\em Phys. Rev. Lett.}, vol.~117, no.~14,
  p.~141801, 2016,
  \href{http://dx.doi.org/10.1103/PhysRevLett.117.141801}{{doi:10.1103/PhysRevLett.117.141801}},
   \href{http://arxiv.org/abs/1602.01086}{{\ttfamily
  arXiv:1602.01086\,[hep-ph]}}.

\end{thebibliography}
\bibliographystyle{arxiv-bibstyle/hyperieeetr2}
%%%%%%%%%%%%%%%%%%%%%%%%%%%%%%%%%%%%%%%%%%%%%%%%%%%%%%%

\end{document}